# INTRODUCING GEOHEAT: GEORADAR-AIDED HIGH-RESOLUTION EXPLORATION FOR ADVANCING GEOTHERMAL ENERGY USAGE

Alexis Shakas[1], Linus Villiger[1], Edoardo Pezzulli[1,2], Matthew Schubert[3], Johan Friborg[4], Anton Nordenstam[4], Arnaud Mignan[5], Paul Lehmann[6], Dieter Werthmüller[1], Maren Breme[1], Michèle Marti[1], Stefan Wiemer[1], Geneviève Savard[7], Francisco Munoz-Burbano[7], Matteo Lupi[7], Christin Bobe[8], Florian Wellmann[8,11], Ezgi Satiroglu[8], Tabea Kautz[8], Kavan Khaledi[11], Francesco Grigoli[9,10], Claudia Finger[11], Erik Saenger[11], Evert Slob[12], Katrin Loer[12]

[1] ETH Zürich, Institute of Geophysics, Sonneggstrasse 5, Zurich, Switzerland

[2] Storra Dynamics GmbH, Habsburgstrasse 35, Zürich, Switzerland

[3] Advanced Logic Technology, 30 H rue de Niederpallen, L-8506 Redange, Grand-Duchy of Luxembourg

[4] Guideline Geo, Hemvarnsgatan 9, Solna, Sweden

[5] Mignan Risk Analytics GmbH, Brunnenstrasse 32, Uster, Switzerland

[6] Bo-Ra-tec GmbH, Hegelstrasse 5, Weimar, Germany

[7] University of Geneva, 13 Rue des Maraîchers, Geneva, Switzerland

[8] RWTH Aachen University, Computational Geoscience, Geothermics, and Reservoir Geophysics, Templergraben 55, Aachen, Germany

[9] University of Pisa, Lungarno Pacinotti 43/44, Pisa, Italy

[10] SEISMIX S.R.L., Via Briuccia 70, 90146 Palermo, Italy

[11] Fraunhofer Society for the Advancement of Applied Research e.V., in Hansastraße 27C, München, Germany

[12] Delft University of Technology, Stevinweg 1, 2628 CN, Delft, Netherlands

email address of (main) author: alexis.shakas@eaps.ethz.ch

**Keywords:** deep, georadar, borehole, passive seismics, techno-economic analysis.

## ABSTRACT

We present a novel geothermal exploration approach that integrates innovations at three spatial scale. At the regional scale (~100 km) we create LCOE heat maps using a techno-economic and metamodel analysis. This allows us to choose several potential sites to perform a reservoir scale (~10 km) assessment with passive seismics and gravity. By integrating the latter with probabilistic geological and geomechanical modeling we propose locations to drill exploration boreholes. Then follows a high-resolution borehole characterization incorporating various analyses, the central one being a georadar probe that allows to illuminate permeable structures. Within this project, we also design and build a geothermal grade georadar. Here, we present the approach and result from the application of our methodology to the Swiss canton of Thurgau, with the purpose of large-scale geothermal exploration.

## 1. INTRODUCTION

Deep geothermal energy harnessed from the subsurface at depths greater than 400 meters could play a vital role in securing a sustainable and independent energy future for Europe. It may offer reliable baseload power that complements intermittent renewable sources, reduces reliance on fossil fuels and nuclear power, and provides heat that can displace fossil-fuel-based systems, and in the meantime limit the expansion of electricity-based heating. However, despite its considerable potential, the widespread adoption of deep geothermal energy is hindered not only by technical challenges but also by difficulties in gaining public and political acceptance (Bressan et al., 2024). These challenges are often exacerbated by poor communication about uncertainties and the risks associated with induced seismicity, which can range from minor disturbances to events severe enough to jeopardize entire projects. Together, these factors contribute to non-competitive Levelized Costs of Electricity (LCOE) and Heat (LCOH), limiting the sector's scalability and impact (Soltani et. al. 2021).

At the early stages of a project - starting from a greenfield site and progressing from the regional scale (~100 km) to the reservoir scale (~10 km) - it is technically challenging to identify productive geothermal reservoirs or to detect hazardous faults before drilling begins, especially at greater depths. The challenge begins with properly analyzing all available data and selecting a location for more focused exploration.

Current pre-drilling exploration approaches are often costly and predominantly deterministic, relying heavily on expensive active seismic campaigns that may not be well-suited for imaging deep geothermal resources (Kana et al., 2015). The latter is particularly true in deep crystalline basements, where the seismic energy from surface-based (active) methods is unlikely to reach target depths. Passive seismic methods like local earthquake tomography and ambient seismic noise offer alternatives but require significant advancements in data processing and interpretation to accurately characterize geothermal resources. Furthermore, typical geomodelling approaches often provide only single realizations of subsurface structures, lacking rigorous frameworks to quantify uncertainty.

Once a reservoir is identified, additional technical challenges arise at the borehole scale, i.e., the tens meters around the borehole. Effectively characterizing this scale is imperative to ensure the establishment, but also the reliable and safe operation of a future geothermal plant. A fundamental bottleneck is the limited imaging reach and capabilities of existing borehole (downhole) tools. Traditional downhole tools have a sensitivity confined to within a few meters of the borehole wall, providing only a limited representation of the subsurface. One technology that offers a promising alternative is georadar, also called Ground Penetrating Radar (or GPR, for short). Georadar, especially when applied to crystalline basement rocks, is perhaps the most promising technology to image permeable structures (Shakas et al, 2020). Despite this, existing probes are not designed for the high-temperature and high-pressure conditions of deep geothermal wells. In addition, georadar only offers a proxy (i.e., the dielectric properties) to the rock properties or real interest, like permeability and porosity of the rock.

To overcome this, data from the borehole must be integrated with georadar images to obtain an informative model of the target reservoir. Digital Rock Physics (DRP) models, arising from analysis of rock cuttings that are a natural byproduct of drilling, can offer insights that link dielectric properties to geomechanical and hydraulic properties (Dvorkin et al., 2013) Traditional wireline logging datasets, in particular acoustic televiewer images, are essential to orient borehole structures and quality-check the DRP models. Real-time monitoring and accurate mapping of drilling-induced micro-seismicity can be combined to this analysis to include seismic risk into structural models. As a result, seismically susceptible fractures and faults could be identified as early as possible in a geothermal project.

The GeoHEAT project, funded by the European Union under Horizon Europe and the SERI, directly addresses these limitations with the ambition to enable a widespread and secure exploitation of deep geothermal energy in Europe. GeoHEAT proposes two strategic shifts: 1) minimizing pre-drilling exploration costs while quantifying and communicating risks and 2) maximizing the learning rates of exploration drilling while reducing characterization costs. The vision is to elevate drilling success rates while reducing LCOE costs by drastically reducing exploration costs.

To achieve this, we propose a novel geothermal exploration workflow that is scalable, low-cost, and rapidly executed. This workflow integrates technical evaluations across regional, reservoir, and borehole scales with crucial social and economic factors. It employs innovative passive geophysical methods and probabilistic modelling for low-cost, uncertainty-encompassing pre-drilling assessment. A pioneering geothermal grade georadar probe, complemented by other advanced techniques, provides high-resolution characterization around the borehole.

In this contribution, we begin by outlining the methodology of our approach. We then present results from a regional and reservoir scale exploration in the Canton of Thurgau, Switzerland, as well as a borehole-scale characterization from the Bedretto Underground Laboratory for Geosciences and Geoenergies. In addition to the current results, we discuss our projects' plans to develop robust reservoir modelling tools and advanced visualization and decision-making tools to enhance public acceptance.

## 2. MULTI-SCALE APPROACH

The GeoHEAT project employs an innovative, multiscale exploration workflow designed to be scalable, low-cost, and rapidly executed, integrating technical, social, and economic factors for successful geothermal projects. This workflow operates across three interconnected scales, which are integrated in the last step of the workflow.

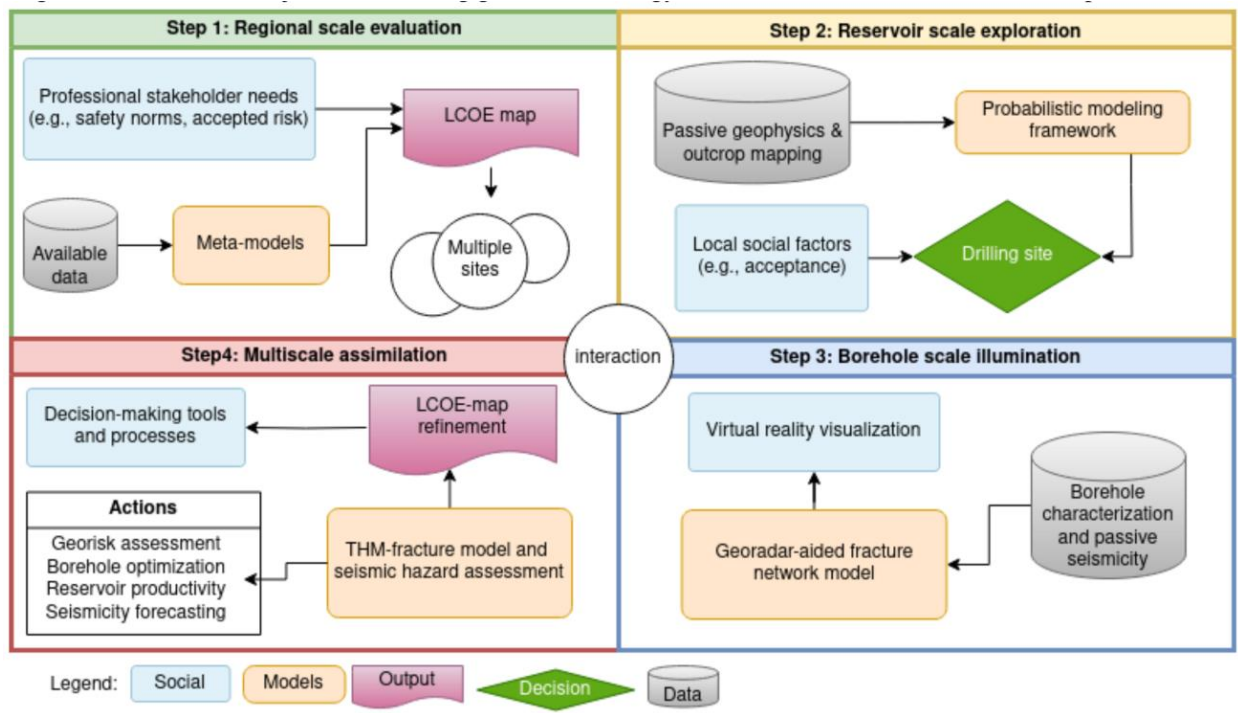

**Figure 1: A schematic presenting the GeoHEAT multi-scale approach, which combines the regional, reservoir and borehole scales with a final multi-scale assessment.**

### 2.1 Regional Scale

A socio-techno-economic evaluation is conducted to identify optimal sites for geothermal exploitation over areas spanning 50-100 km. This involves integrating existing data on seismic risk, geothermal productivity, and public acceptance to create an LCOE-heat-credit map that highlights promising locations while considering social and environmental factors. In addition, an initial, coarse regional-scale passive seismic survey will be conducted, and its results will be integrated into the regional-scale assessment, identifying multiple sites suitable for more detailed exploration.

### 2.2 Reservoir Scale

At this scale (approximately 10 to 20 km), low-cost and time-efficient passive seismic and gravity campaigns are performed at multiple potential sites identified in the regional assessment. In this phase, geological information from regional models is integrated with local geological data to build a comprehensive geological model of the site. Subsequently, geophysical and geological information are integrated within a probabilistic modeling workflow that quantifies structural uncertainties in the geological model. Geomechanical simulations are then performed to screen fault stability issues and select optimal exploration borehole locations.

**2.3 Borehole Scale**

Once a borehole is drilled, a high-resolution characterization of the surrounding geothermal reservoir is performed. This involves deploying borehole seismometers and surface stations to monitor micro-seismicity during drilling (Shi et.al., 2022), analysing drill cuttings with digital rock physics to characterize rock properties, and utilizing a novel geothermal-grade georadar probe and acoustic televiewer to create a high-resolution 3D fracture network model up to 100m away from the borehole.

**2.4 Multi-scale integration**

The final phase of our analysis will include a multi-scale integration of the results from the previous 3 steps. The aim here is to integrate the high-resolution borehole characterization into an encompassing Thermo-Hydro-Mechanical (THM) model of the reservoir to improve assessments of productivity and optimise exploitation strategies. The imaged 3D fracture networks are upscaled through Thermo-Hydraulic simulations of explicitly represented 3D fracture networks, capturing the anisotropic permeability and thermal conductivity relevant to fractured and faulted geothermal systems. At the same time, the regional-scale information (e.g., on stress regimes, large-scale geology and overburden) informs geomechanical boundary conditions and large-scale structural features, which is combined with constraints obtained from drilling-induced micro-seismicity and mud-losses on fault stability and fault conductivity. Overall, the multi-scale understanding aims to deliver insights on optimal borehole deviations to target conductive faults and fault-intersections, assist in designing hydraulic stimulations informed by seismic risk and fault slip tendencies, optimally place secondary production and/or injection wells, and feed back into improved regional assessments of geothermal resources.

## 3. RESULTS

The results presented in this section reflect the initial findings from the first 12 months of the GeoHEAT project, which began in June 2024 and runs for a total duration of four years. At this stage, the work has focused on two key areas: i) regional-scale exploration in the canton of Thurgau, and preliminary social science investigations into public and stakeholder perspectives on geothermal development in the region, and ii) the first-generation design and prototyping of a georadar antenna for borehole-scale investigations.

**3.1 Regional Scale Exploration**

The regional scale exploration for our project is currently taking place in the canton of Thurgau in collaboration with [Geothermie Thurgau AG](#), and support from the cantonal authorities. The exploration started with a two-fold analysis of all available data at the entire cantonal scale and proceeded with a deployment of passive seismic sensors across the whole canton.

**Techno-Economical Analysis**

The techno-economic assessment of the geothermal potential in the Canton of Thurgau was, in a first step, based on a spatially differentiated cost model, including drilling costs, operational expenditures (OPEX) and various infrastructure scenarios, for three target temperature levels: 60°C, 100°C and 150°C. In addition to economic aspects, the analysis considered regions with high heat demand and areas with generally favourable geological conditions. The calculated total costs were compared with geologically suitable zones and areas of high heat demand to provide a realistic evaluation of location potential. A key finding of the study is the particularly high suitability of southern Thurgau for geothermal development. This refers to the corridor extending from Frauenfeld southwards to the municipalities of Aadorf, Münchwilen and Eschlikon, which together form a coherent region with favourable geological and economic conditions. These areas were identified as especially promising based on geological formations, demand patterns and existing infrastructure. This can be seen in Figure 2, that outlines the total costs when utilizing water at 60°C for heat generation, including any additional infrastructure components.

Another central result concerns the role of infrastructure costs. Contrary to initial expectations, these do not represent the dominant cost factor in the overall model. This is largely due to the well-developed road network in the Canton of Thurgau. Instead, operational expenditures account for a significant share of total costs – especially at higher target temperatures. The analysis also shows that connecting to existing district heating networks is more cost-effective than establishing new electricity transmission lines.

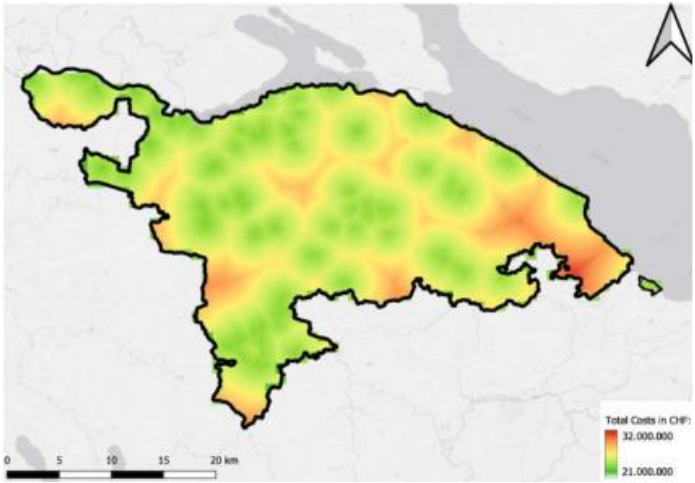

**Figure 2: Total costs (in CHF) for geothermal utilisation at 60°C in the canton of Thurgau, including additional infrastructure components such as roads and district heating networks. The limits range from 21M CHF to 32M CHF.**

**Meta-models and LCOE optimization**

Favourability maps were also generated through the application of the Levelized Cost of Electricity (LCOE) meta-model of Mignan et al. (2019). Dozens of parameters - both generic and region-specific - were carefully defined to assess the economic feasibility of geothermal energy projects in the Thurgau area. LCOE maps were explored under various parameterizations, including standalone LCOE, LCOE with heat credit considerations, and scenarios incorporating the costs of seismic risk mitigation. These latter cases were analysed under both neutral and risk-averse perspectives. The results provide valuable insights into the economic potential of geothermal energy in Thurgau, supporting informed decision-making for the next stages of the GeoHEAT project. Results will be refined toward the end of the project once more local data are available with the goal to reduce the LCOE based on the various findings of the project.

Figure 3 shows some preliminary results for the default parameterisation considered in GeoHEAT, here for a borehole depth of 6 km. The top left map first shows the geothermal map of the Thurgau region, which indicates higher temperatures in the southern area. This is reflected in the standard LCOE map. Minimum, median and maximum values are 0.17, 0.27 and 0.55 CHF/kWh. These values can be significantly reduced by the heat credit, with 0.09, 0.22 and 0.48 CHF/kWh. Due to the importance of the heat credit, optimal areas to site a geothermal plant become areas near the largest population centres. This shows the importance of considering the impact of seismic risk mitigation measures. The probability of failing a standard safety norm of $10^{-6}$ probability of a statistical death is found to be 23% on average. The impact of these measures is not so visible from the new LCOE estimates in the risk neutral case with the overall costs increased by 4% on average. However, in the risk averse case, all values significantly increase with 0.13, 0.31 and 0.75 CHF/kWh obtained. These results remain subject to high epistemic uncertainties and should be used as proof of concept and not a map for direct commercialisation considerations at this stage.

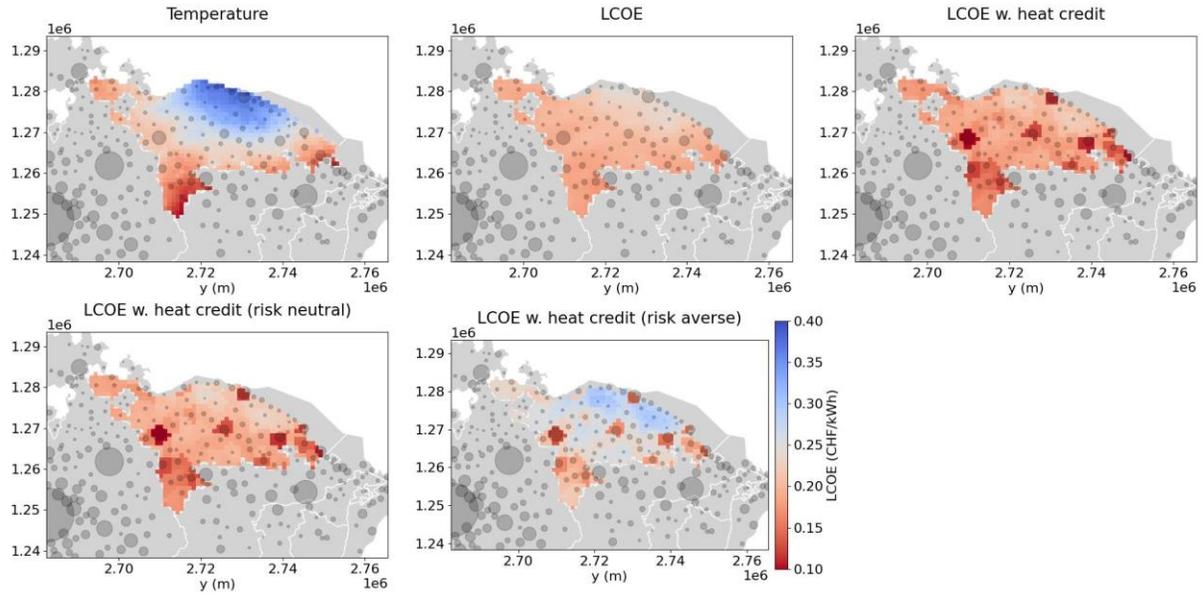

**Figure 3: LCOE maps for standard parameterisation, shown for a borehole at a depth of 6 km.**

**Passive Seismic campaign**

From March to April 2025, we performed a passive seismic campaign in the Canton of Thurgau where we deployed 299 nodal 5 Hz geophones (SmartSolo IGU-16HR 3C) that acquired data for a period of roughly 1 month. The deployment map is shown in Figure 4. The deployment took 12 people 5 days to complete, and a smaller team recovered the nodes in 3 days. The data processing and inverse modelling are still in progress, but early results are encouraging. Although no local earthquakes occurred during the deployment, high-quality noise cross-correlation functions have revealed clear surface-wave propagation.

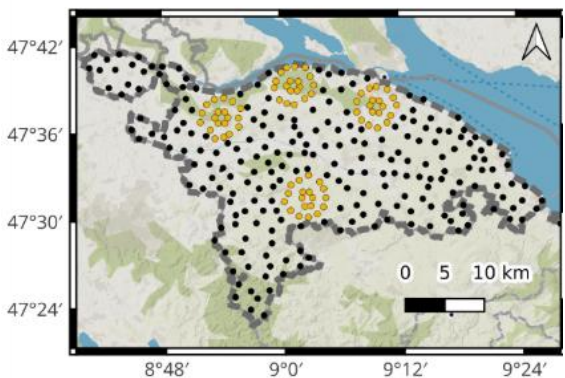

**Figure 4: 299-node seismic network deployed in Thurgau covering 990 km². The circular-like arrays (highlighted in yellow) correspond to the four sub-array called "antennas".**

By inverting group-velocity measurements at multiple frequencies after NANT (Savard et al., 2023), we have recovered key geological features from group velocity maps (see examples in Figure 5) - most notably the eastern terminus of the northern Swiss Carboniferous Trough (the "Constance–Frick trough") and smaller troughs in the canton's southeast and southwest corners. At shallower depths, the resulting maps clearly delineate deep sedimentary channels and hint at a possible deep groundwater reservoir along the shores of Lake Constance. Deep basement structures - such as the Permo-Carboniferous troughs - are important exploration targets because they exert primary control on fluid and heat flow at depth and govern the distribution of fracture zones. Future processing steps will allow us to better constrain their geometry.

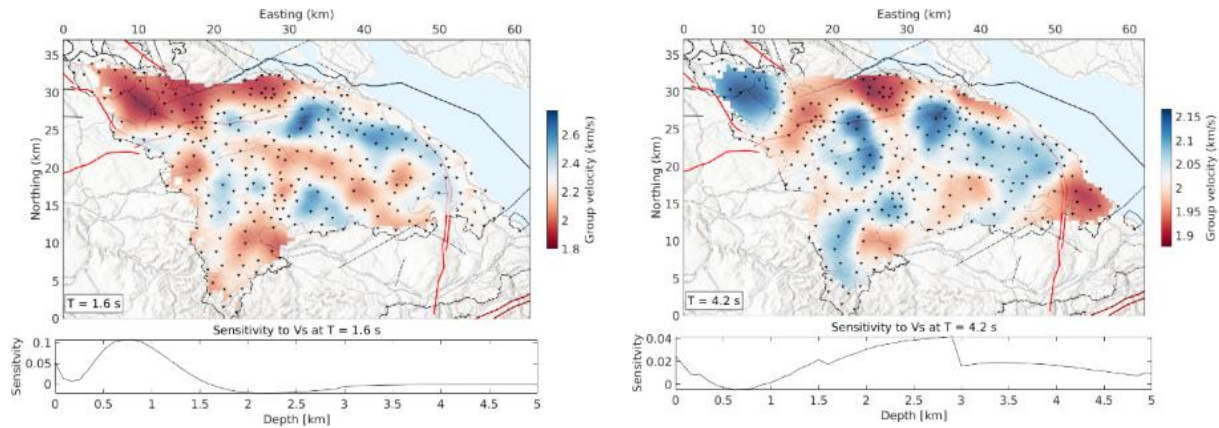

**Figure 5: Group-velocity maps at periods of 1.6 s (top) and 4.2 s (bottom), highlighting the northern Permo-Carboniferous trough in the canton's northwestern corner, a deep SE–NW sedimentary channel across the study area, and smaller troughs along the southern border. Major surface-fault traces from the Swisstopo database are overlaid in red. Below, the corresponding estimated depth-sensitivity kernels for shear-wave velocity at each period are shown, based on well-log data and previous seismic studies.**

### 3.2 Reservoir scale exploration

Based on the insights gained from the previous regional scale analysis, specific sites for more detailed investigation will be selected and further examined. We will achieve this by integrating the regional scale analysis with local geological cross-sections, leading to the construction of a three-dimensional geological model. Uncertainties in the input geological maps and geophysical modeling results will allow us to frame the model construction in a probabilistic framework.

### 3.3 Borehole scale exploration

A critical component for extending the borehole scale exploration is the development of the novel geothermal grade georadar instrument. The technical challenges faced during this design include controlling high temperatures, high pressures, corrosive environments, and successfully communicating with the system when deployed at depths of several kilometres.

To date, the system design of a wireline deployed low temperature georadar tool has been completed (Figure 6). Work is progressing on the detailed electronic, mechanical and firmware design for this instrument, which will be used as a testbed to validate the performance of the design before moving on to building a geothermal grade instrument. In parallel, testing has begun on antenna performance using a prototype antenna in the configuration outlined in Figure 6.

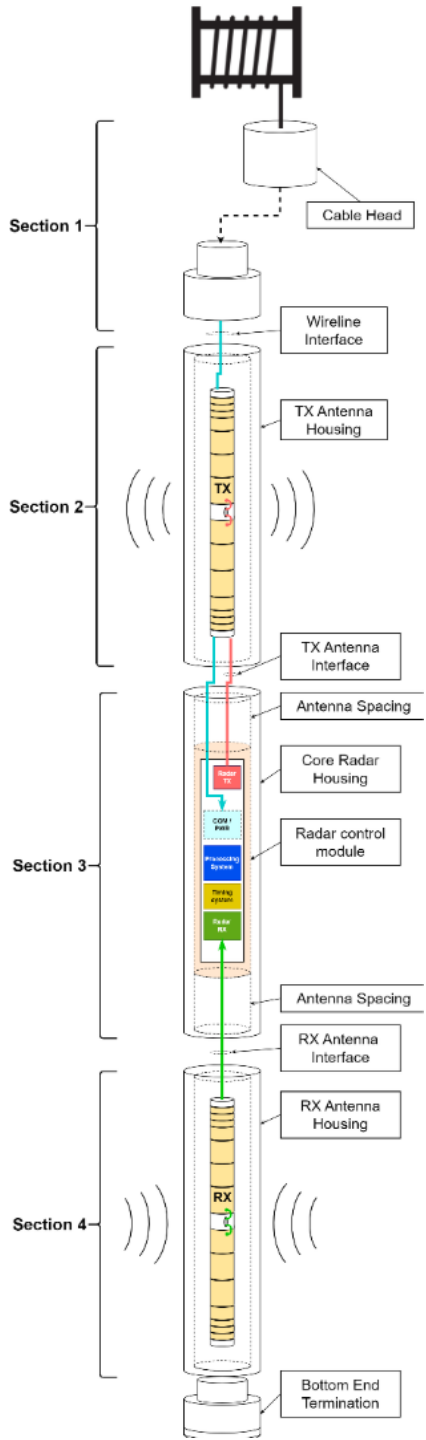

**Figure 6: First-generation design of the georadar antenna for geothermal exploration. The design is separated in 4 sections; section 1 allows communication to the surface, section 2 consists primarily of the transmitter antenna, section 3 hosts the core modules for creating, recording and digitizing the signal, and section 4 is the receiver antenna.**

Another major component is the environmental protection hardware (pressure and thermal barriers) which will allow the tool to operate in a geothermal well. The instrument is required to have a non-conductive pressure housing over the antenna sections, as well as a vacuum flask and heatsinking of the electronics to allow them to survive in the high temperature environment. The system design will include detailed calculations and material specifications for the protection system. We are currently evaluating material choices and drafting the design schematics for the geothermal-grade tool.

## 4. CONCLUSIONS

We introduce a novel approach to geothermal exploration that we will further develop through the GeoHEAT project. The exploration follows a top-down approach, starting with all available information at the regional scale (~100 km) and progressively integrating data and models to choose potential sites for exploration. The chosen sites become new exploration targets at the reservoir scale, which ranges in size between 10 to 20 km. Using passive geophysical methods and probabilistic geological modelling we rank the potential sites based on their geothermal prospect, as well as all factors that minimize LCOE (seismic risk, proximity to infrastructure, feasibility of drilling, amongst others). At the next step we highlight the importance of drilling exploration boreholes (i.e., the borehole scale). Boreholes are financially feasible in our exploration approach without drastically elevating LCOE because of the reduced surface deployment costs of passive geophysical methods. To leverage the use of boreholes, we develop a geothermal grade georadar tool that can image structures up to 100 m away from the borehole wall in crystalline basement conditions. The georadar development is complemented by further innovations such as rock property estimation from cuttings, advanced georadar modelling and detection of drilling-induced seismicity, which lead to a high-resolution structure delineation. Finally, the analysis from the three scales will be integrated into a coupled Thermo-Hydro-Mechanical model which will facilitate the design and guide the exploitation of deep geothermal targets. Here, we presented novel analyses of the regional scale at the canton of Thurgau, advancements in the prototype georadar tool and outlined the planned activities for the upcoming year. Results from the regional and borehole scales will be available within this year.

**ACKNOWLEDGEMENT**

This work has received funding from the European Union's Horizon Europe research and innovation program for the project GeoHEAT (Advanced exploration technologies for geothermal resources in a wide range of geological settings - HORIZON-CL5-2023-D3-02-05). Swiss partners are supported by the Swiss State Secretariat for Education, Research and Innovation (SERI). The authors would also like to thank the Geothermie Thurgau AG and Bernd Frieg for their invaluable support in facilitating access to the regional-scale study site in the canton of Thurgau, as well as for their valuable contributions to discussions on regional-scale assessments and social aspects of GeoHEAT.